\documentclass[twocolumn]{aastex631}

\usepackage{amsmath}

\begin{document}

\newcommand{\nh}{$N_H$}
\newcommand{\source}{MAXI J1807$+$132}
\newcommand{\nicer}{\textit{NICER}}
\newcommand{\swift}{\textit{Swift}}
\newcommand{\dbb}{\texttt{diskbb}}
\newcommand{\bb}{\texttt{bbodyrad}}
\newcommand{\nthc}{\texttt{nthComp}}
\newcommand{\tbabs}{\texttt{TBabs}}
\newcommand{\tbfeo}{\texttt{TBfeo}}
\newcommand{\tbvarabs}{\texttt{TBvarabs}}
\newcommand{\chisq}{$\chi^{2}$}
\newcommand{\redchisq}{$\chi^{2}_{\nu}$}
\newcommand{\kte}{$kT_e$}
\newcommand{\inptype}{$inp_{type}$}
\newcommand{\tin}{$T_{in}$}
\newcommand{\kt}{$kT$}
\newcommand{\ktbb}{$kT_{bb}$}
\newcommand{\gama}{$\Gamma$}
\newcommand{\norm}{\texttt{norm}}
\newcommand{\gp}{$g^\prime$}
\newcommand{\rp}{$r^\prime$}
\newcommand{\ip}{$i^\prime$}
\newcommand{\zs}{$zs$}
\newcommand{\fluxcgs}{erg s$^{-1}$ cm$^{-2}$}
\newcommand{\numax}{$\nu_{max}$}

\graphicspath{{./}{./figures/}}

\title{Evolution of the Accretion Disk and Corona During the Outburst of the Neutron Star Transient MAXI J1807$+$132.}

\author[0000-0001-7590-5099]{Sandeep K. Rout}
\affiliation{New York University Abu Dhabi, PO Box 129188, Abu Dhabi, UAE}
\affiliation{Center for Astrophysics and Space Science (CASS), New York University Abu Dhabi, UAE}

\author[0000-0002-3348-4035]{Teo Mu\~{n}oz-Darias}
\affiliation{Instituto de Astrof\'{i}sica de Canarias (IAC), V\'{i}a L\'{a}ctea s/n, La Laguna E-38205, S/C de Tenerife, Spain}
\affiliation{Departamento de Astrof\'{i}sica, Universidad de La Laguna, La Laguna E-38205, S/C de Tenerife, Spain}

\author[0000-0001-8371-2713]{Jeroen Homan}
\affiliation{Eureka Scientific, Inc., 2452 Delmer Street, Oakland, CA 94602, USA}

\author[0000-0002-4344-7334]{Montserrat Armas Padilla}
\affiliation{Instituto de Astrof\'{i}sica de Canarias (IAC), V\'{i}a L\'{a}ctea s/n, La Laguna E-38205, S/C de Tenerife, Spain}
\affiliation{Departamento de Astrof\'{i}sica, Universidad de La Laguna, La Laguna E-38205, S/C de Tenerife, Spain}

\author[0000-0002-3500-631X]{David M. Russell}
\affiliation{New York University Abu Dhabi, PO Box 129188, Abu Dhabi, UAE}
\affiliation{Center for Astrophysics and Space Science (CASS), New York University Abu Dhabi, UAE}

\author[0000-0003-0168-9906]{Kevin Alabarta}
\affiliation{New York University Abu Dhabi, PO Box 129188, Abu Dhabi, UAE}
\affiliation{Center for Astrophysics and Space Science (CASS), New York University Abu Dhabi, UAE}

\author[0000-0002-5319-6620]{Payaswini Saikia}
\affiliation{New York University Abu Dhabi, PO Box 129188, Abu Dhabi, UAE}
\affiliation{Center for Astrophysics and Space Science (CASS), New York University Abu Dhabi, UAE}

\correspondingauthor{Sandeep K. Rout}
\email{sandeep.rout@nyu.edu}

%% check the luminosity, compare with literature (including radius)

\begin{abstract}

Low-mass X-ray binaries  with a neutron star as the primary object show a complex array of phenomenology during outbursts. The observed variability in X-ray emission primarily arises from changes in the innermost regions of the accretion disk, neutron star surface, and corona. In this work, we present the results of a comprehensive X-ray spectral and timing analysis of the neutron star transient MAXI J1807$+$132 during its 2023 outburst using data from the NICER observatory. The outburst is marked by a very rapid rise in the count rate by about a factor of 20 in a day. The source undergoes full state transitions and displays hysteresis effect in the hardness and rms intensity diagrams. Spectral analysis with a three-component model is consistent with disk truncation during the hard states and reaching the last stable orbit during the intermediate and soft states. We discuss the different values of the last stable radius in the context of possible distance of the source and magnetic field strength. The characteristic frequencies throughout the hard and intermediate states are found to be strongly correlated with the inner radius of the disk. Together with the spectral and fast variability properties, we attempt to trace the evolution of the size of the corona along the outburst. Following the main outburst, the source undergoes a high amplitude reflare wherein it shows a complex behavior with relatively high variability (10$\%$), but low hardness.

\end{abstract}


\section{Introduction} \label{sec:intro}

The deaths of massive stars ($\gtrsim 8 M_\odot$) result in the formation of compact objects such as neutron stars (NSs) and black holes (BHs). When these compact stars occur in a binary system with a late-type or main sequence companion, they accrete matter during certain stages of the binary evolution. The mode of accretion depends on the size of the companion and/or the orbital parameters of the binary \citep[e.g.][]{frank02}. As matter falls into the deep gravitational wells of the compact objects, it radiates copious amounts of energy. While the frequency of this radiation spans the entire electromagnetic spectrum, the majority of the emission is in the form of X-rays. The primary emitting component of an X-ray binary (XRB) is an optically thick and geometrically thin accretion disk extending from close to the NS/BH to near the radius of the companion. The inner regions of the disk are often surrounded by a hot ($\sim 10^{7-9}$ K) cloud of electrons, often referred to as the corona \citep[for reviews on the topic, see][and references therein]{remillard06,done07,belloni16}. Apart from matter accreting onto the compact objects, in most XRBs, matter also flows out of the system as jets and winds. These outflows can channel energy, mass, and angular momentum out of the system and form an integral part of the accretion-ejection paradigm in XRBs \citep[see][and references therein]{fender16}. The interplay of all these structures and their associated phenomena leave a characteristic imprint on the observed emission. Deciphering the nature, origin, and geometry of the structures, as well as their temporal evolution, from observations is challenging and a matter of debate. 

Low-mass X-ray binaries (LMXBs) are mostly transient objects wherein they undergo periods of enhanced mass accretion, or outbursts, spanning a few weeks to several months, interspersed between much longer periods of inactivity or quiescence \citep[see][for a recent review]{bahramian23}. Throughout an outburst, LMXBs pass through various phases, or states, characterized by different spectral and timing properties. On an X-ray hardness-intensity diagram \footnote{A plot between X-ray hardness, defined as the ratio of counts in hard to that in soft energy bands, versus total luminosity.} (HID) they trace a hysteresis curve in the shape of a `q' \citep[e.g.][]{homan01,maccarone03, homan05b}. This implies that the transition from hard (typically characterized by high hardness and variability) to soft (low hardness and variability) state during the rise occurs at a much higher luminosity than the reverse transition from soft to hard states during the decay \citep[e.g.,][]{munozdarias14}. 

Owing to the presence of a hard surface, which anchors strong magnetic fields, LMXBs with NS as the accretor show somewhat different and additional phenomenology which are often used in identifying them. For instance when the material accreted onto a NS surface reaches a critical density a runaway thermonuclear explosion takes place which is classified as a Type-I thermonuclear burst. It appears as a sudden rise in X-ray luminosity over a timescale of seconds \citep{galloway08, galloway20}. Some NS LMXBs also show pulsations in a wide waveband when their magnetic poles are misaligned with their spin axis and the radiation from the poles sweeps the direction of the earth with a period of the NS's rotation. These can be detected both during the bursts, or persistent phases \citep{vanderklis97}. Additionally, NS binaries have been exclusively found to exhibit quasi-periodic oscillations in the kilo-Hertz frequency ($300-1200$ Hz) range, the origin of which is debated \citep[e.g.][]{strohmayer96,lamb01}.

NS LMXBs are classified into `Z' and `atoll' sources based on the shape of the tracks on a color-color diagram \citep[CCD;][]{hasinger89}. A CCD is a representation of the brightness of the source in two different energy bands, usually a hard (e.g. $6-10/4-6$ keV) and soft ($2-4/0.5-2$ keV) color. Z sources are brighter ($\gtrsim 0.5 L_{Edd}$, where $L_{Edd}$ is the Eddington luminosity) than atoll sources ($0.001 - 0.5 L_{Edd}$) and have different spectral and timing properties based on their position on the CCD \citep{vanderklis06}. The two major states of atolls are island and banana branches which are further subdivided into extreme island, island, and left-lower, lower, and upper banana branches. In the island branches, the source is often in a low luminosity hard state and shows high variability. In the banana branches, on the other hand, the source remains soft over a wider range of luminosities and the power spectrum is dominated by a low frequency noise. While the source moves quite fast, over a time scale of hours to a day, within a banana branch, the movement in the island branch is over a time scale of days to a week. This slow motion is attributed to the reason behind `island'-like patches in the CCD owing to gaps in observations \citep[see][for a recent review]{disalvo23}.

While the overall phenomenologies of the various states are well understood, the underlying physical factors that drive them are still uncertain. Of particular interest are the geometrical changes that take place in the inner regions of the accretion disk when the source transitions between the hard and soft states. X-ray spectral and timing variability, with improved instrumentation, serves as a useful tool in this regard. The spectral decomposition of a NS LMXB is more complicated than its BH counterparts, owing to the presence of a solid surface. In weakly magnetic NSs, the accretion disk can extend close to the NS surface, leading to the formation of a shearing sheet of matter known as the boundary layer (BL) \citep{shakura88, inogamov99}. At this layer, the sub-relativistic Keplerian velocity of the inner-disk material decelerates down to the NS's rotation velocity, consequently releasing half of its gravitational energy \citep{gilfanov14}. The BL manifests as a luminous blackbody peaking at soft X-ray energies. Therefore, the spectra of NS LMXBs require two thermal components, one for the BL and the other for the disk emission. These thermal emissions often undergo inverse comptonization in a hot cloud of electrons ($\sim 10-100$ keV) manifesting as a powerlaw in the spectrum with a cutoff at the temperature of the electrons \citep[e.g.][]{lin07,lin09, armaspadilla17}. A fraction of the comptonized photons sometimes scatter back and impinge the disk leading to a reflection component in the spectrum \citep{fabian89}. The reflection spectrum is characterized by a compton hump in $10-30$ keV and a fluorescent K$\alpha$ line from ionized elements, primarily Fe \citep[e.g.][]{asai00}.

\begin{figure}
    \centering
    \includegraphics[scale=0.6]{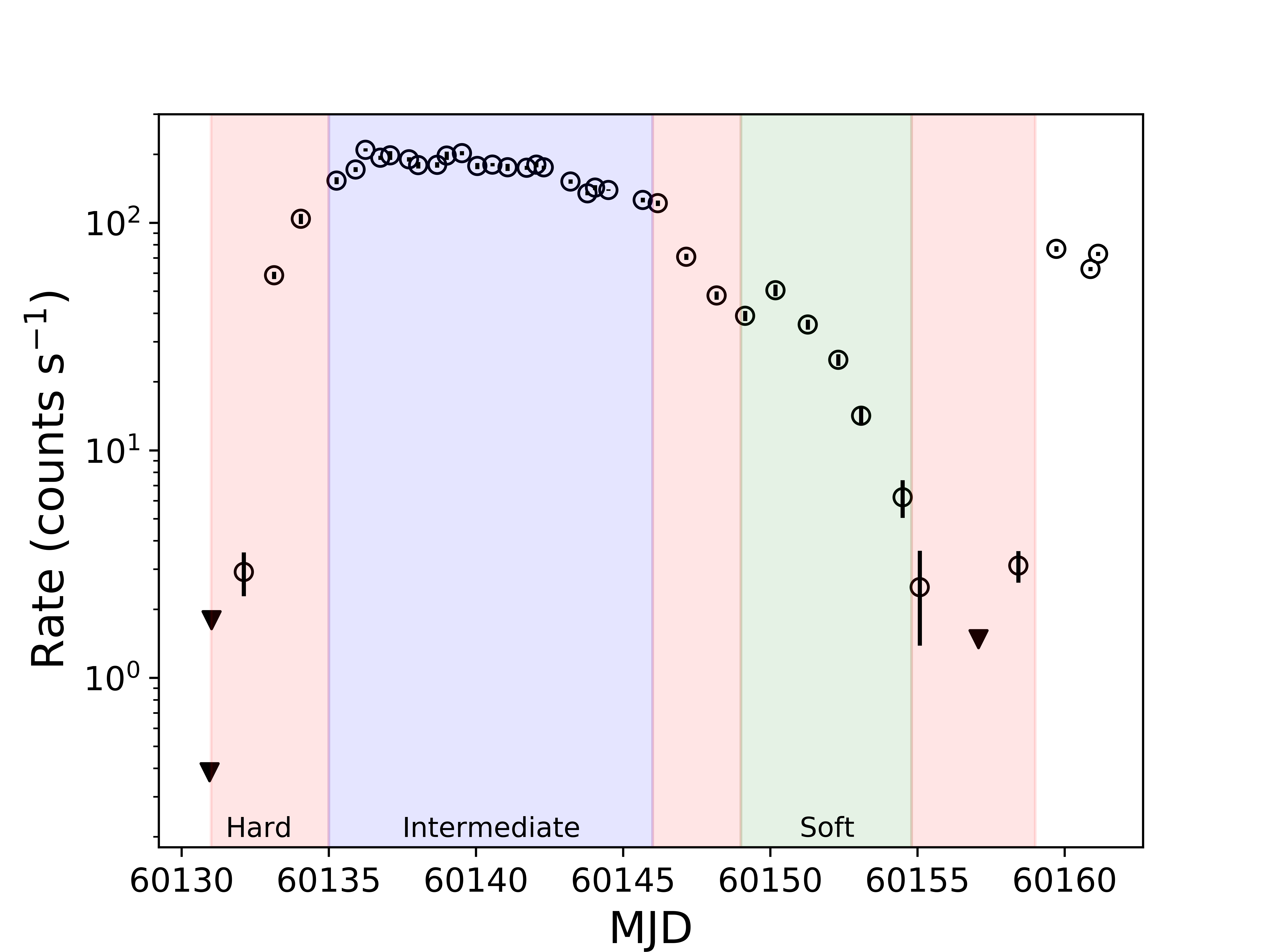}
    \caption{The \nicer{} lightcurve of \source{} containing all observations in the $1-10$ keV range. The points marked with triangles represent upper limits for non-detections, while all the rest are significant detections. The different states along the outburst are marked by different colors.}
    \label{fig:lc}
\end{figure}

Time variability studies have also played a vital role in attempts to constrain the geometry of the disk and corona \citep[e.g.][]{ingram11,karpouzas20}. Rapid variability, produced from the hard X-ray emitting region close to the compact object, is primarily studied through power spectral analyses \citep[e.g.][]{vanderklis97,nowak99}. The characteristic frequencies in the power spectra, especially those of the quasi-periodic oscillations, are known to act as proxies for different length scales such as the size of the corona, the truncation radius of the disk, the position of the transition layer, etc \citep[see e.g.][for reviews on X-ray timing]{belloni02,ingram19b}. 

\source{} was discovered in March 2017 by the Monitor of All-sky X-ray Image \citep[][]{shidatsu17, jiminezibarra19}. It is known to harbour a NS due to the detection of type-I thermonuclear bursts during the 2019 outburst \citep{albayati21}. The source underwent its latest outburst in July 2023 and was observed by several space and ground based observatories \citep{atelilliano23,atelsaikia23a}. In this work, we present a comprehensive X-ray spectral and timing analysis using data from the \nicer{} observatory.

\section{Observations and Data Reduction} \label{sec:obs}

We analyzed all the data acquired between 05 July 2023 to 05 August 2023 by the \nicer{} observatory \citep{gendreau12} for exploring the X-ray properties of \source{}. The analysis was carried out using \texttt{NICERDAS\_V011a} which came with the general purpose X-ray analysis software \texttt{heasoft-6.32.1}. The cleaned event files were generated using the \texttt{nicerl2} module, which was then used for higher level products. In May 2023 a few thermal filters of \nicer{} got damaged resulting in optical light leakage during orbital day \footnote{\url{https://heasarc.gsfc.nasa.gov/docs/nicer/analysis_threads/light-leak-analysis/}}. In order to avoid any spurious signal in our data, we only used data acquired during orbital night (ensured by setting \texttt{thresh\_range} between $-3$-3 in \texttt{nicerl2}). The lightcurves and spectra were extracted with the new modules \texttt{nicerl3-lc} and \texttt{nicerl3-spect}, respectively. For the faint phases of the outburst during the rapid rise and decay, where the analysis was done in smaller segments, we used the \texttt{SCORPEON} background model \citep{markwardt24} and for rest of the bright phases we used the \texttt{3C50} model \citep{remillard22}. We ensured that the results of the spectral fits from using the two background models are consistent within uncertainty.  
%% caldb version xti20221001

We studied the rapid variability properties of \source{} using Fourier analysis techniques \citep{vanderklis89b}. The power spectra were extracted with the \texttt{GHATS} \footnote{\url{http://astrosat-ssc.iucaa.in/uploads/ghats_home.html}} package. In order to construct the power spectrum, we averaged over the Fourier transforms of smaller segments ($\sim 26$ seconds) of lightcurves in each observation. This allowed us a frequency resolution and minimum frequency of $\sim 0.03$ Hz. We binned the \nicer{} events by a factor of $10^4$ to achieve a Nyquist frequency of 1250 Hz. After ensuring that the higher frequencies were dominated by Poisson noise and lacked any kHz quasi-periodic oscillations, we subtracted the average power above 200 Hz from the power spectrum and normalized it to the squared fractional rms \citep{belloni90}. 

\begin{figure*}
    \centering
    \includegraphics[scale=.55]{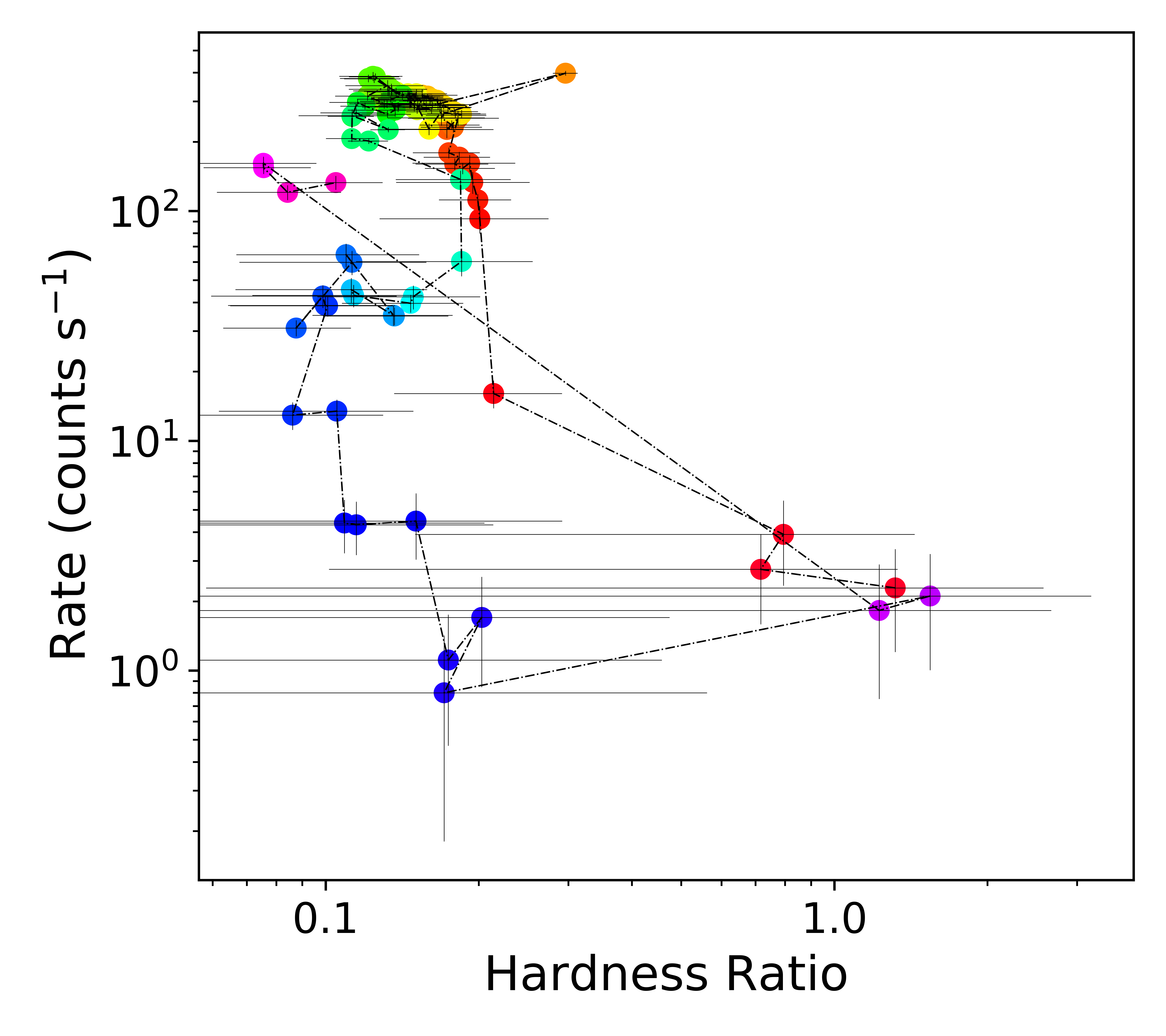}
    \includegraphics[scale=.55]{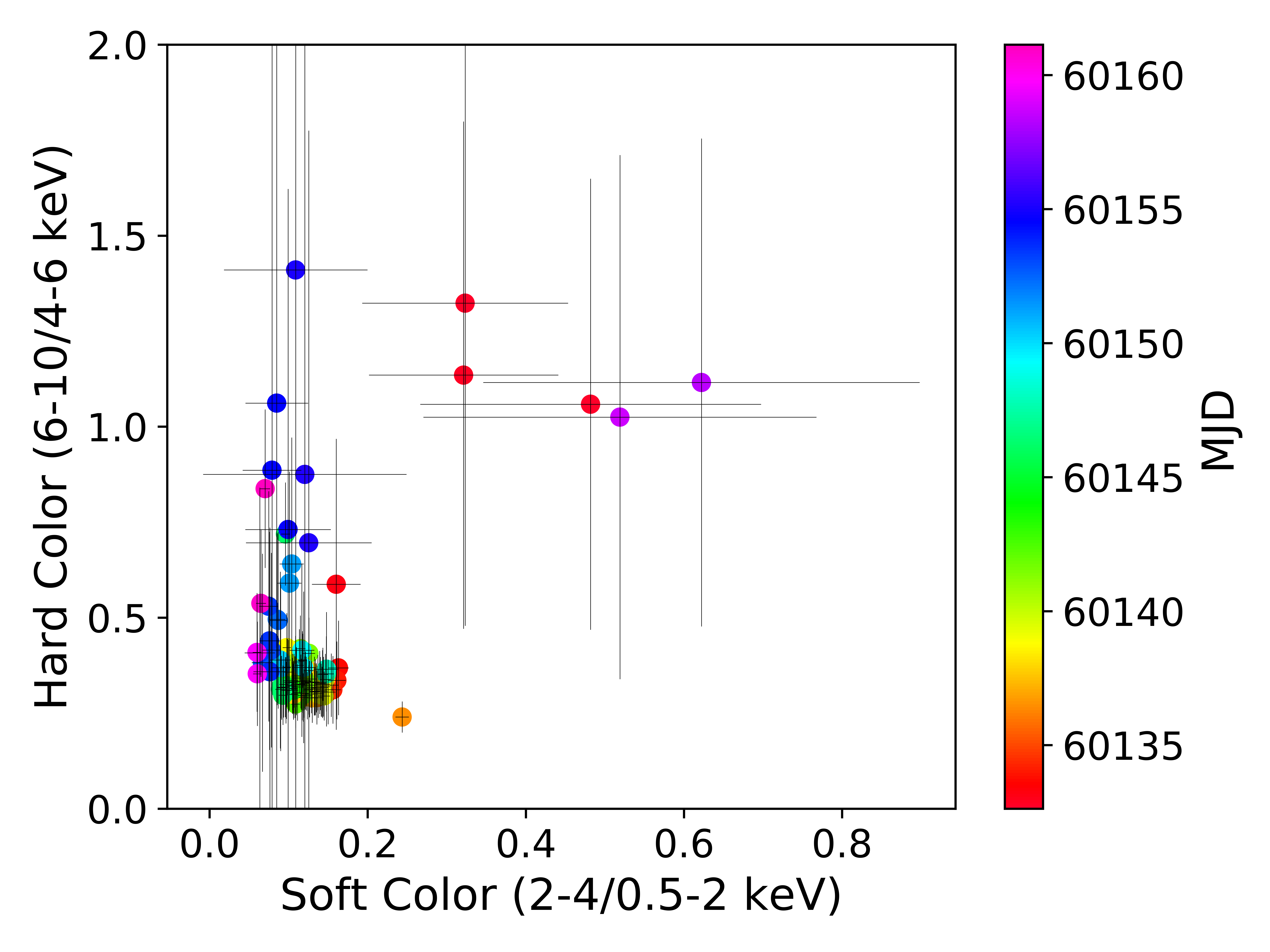}
    \caption{Left: Hardness-intensity diagram (HID) of \source{} during the 2023 outburst. The hardness ratio is defined as the ratio of count rates in $2-10$ keV to $0.5-2$ keV bands, and the total count rate is in the $0.5-10$ keV range. Right: Color-color diagram (CCD) of the source during the same period. The source is in the island branch during the beginning and end of the outburst, and in the banana branch during rest of the outburst. The points are color coded according to the dates of the observations.}
    \label{fig:hidccd}
\end{figure*}

\section{Analysis and Results} \label{sec:analysis}

\subsection{Outburst Evolution}

\nicer{} started observing \source{} from 05 July 2023 (MJD 60130) following the trigger from an optical detection \citep{atelsaikia23a}. In the initial two observations, however, the source brightness was consistent with the background level \footnote{Due to this the two observations on MJD 60130.95 (ObsId: 6200840101) and MJD 60131.02 (ObsId: 6200840102) and the one at the end of the main outburst on MJD 60157.1 (ObsID: 6634010122) have not been analyzed in this work.}. The first significant detection was made on 07 July 2023 (MJD 60132.1) with a count rate of $\sim 3$ counts s$^{-1}$ in the $1-10$ keV range range (Figure \ref{fig:lc}). The count rate rose rapidly in the next few days to attain a local maximum on MJD 60136.2 (ObsId: 6200840107) with $\sim 209$ counts s$^{-1}$. This day also marked the detection of a thermonuclear burst. The period from MJD 60136 to 60146 resembles a ``plateau'' phase wherein the count rate declined very slowly.  
Following this there was a sudden drop in the count rate to $\sim 39$ counts s$^{-1}$ in less than three days, accompanied by an increase in hardness. This culminated in a small flare on MJD 60150.2 when the rate increased slightly to $\sim 50$ counts s$^{-1}$. The count rate continued to decay until MJD 60155.1 to the levels at which it was first detected. On MJD 60158.4 the source seemed to rise back from a background dominated stage to $\sim 3$ counts s$^{-1}$. Observations made on the following three days confirmed a very bright flare with count rate rising to $\sim 70$ counts s$^{-1}$ (Figure \ref{fig:lc}). The lack of follow-up observations made it impossible to characterize the evolution of the flare, including its recovery to quiescent level.

\begin{figure*}
    \centering
    \includegraphics[scale=.55]{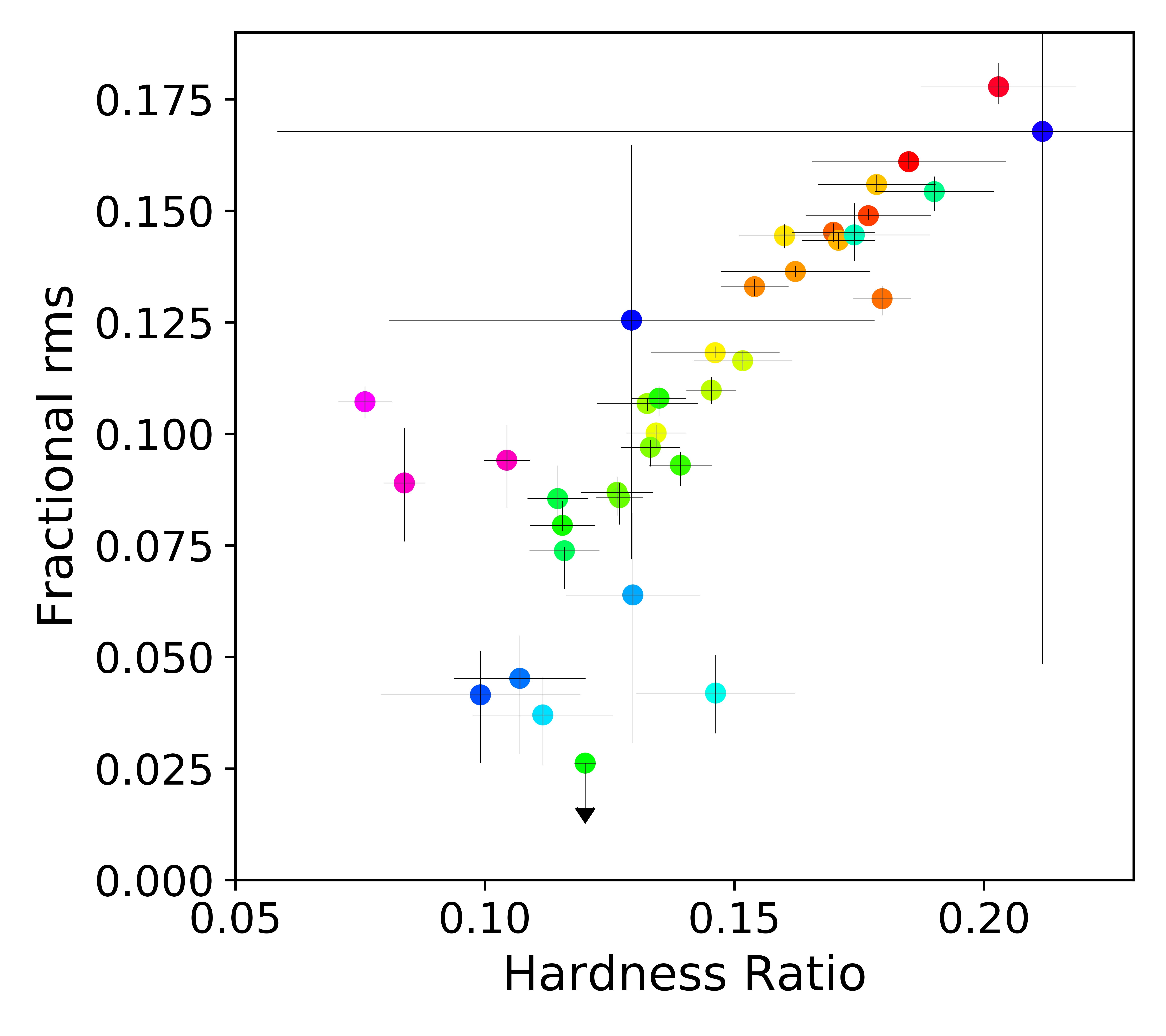}
    \includegraphics[scale=.55]{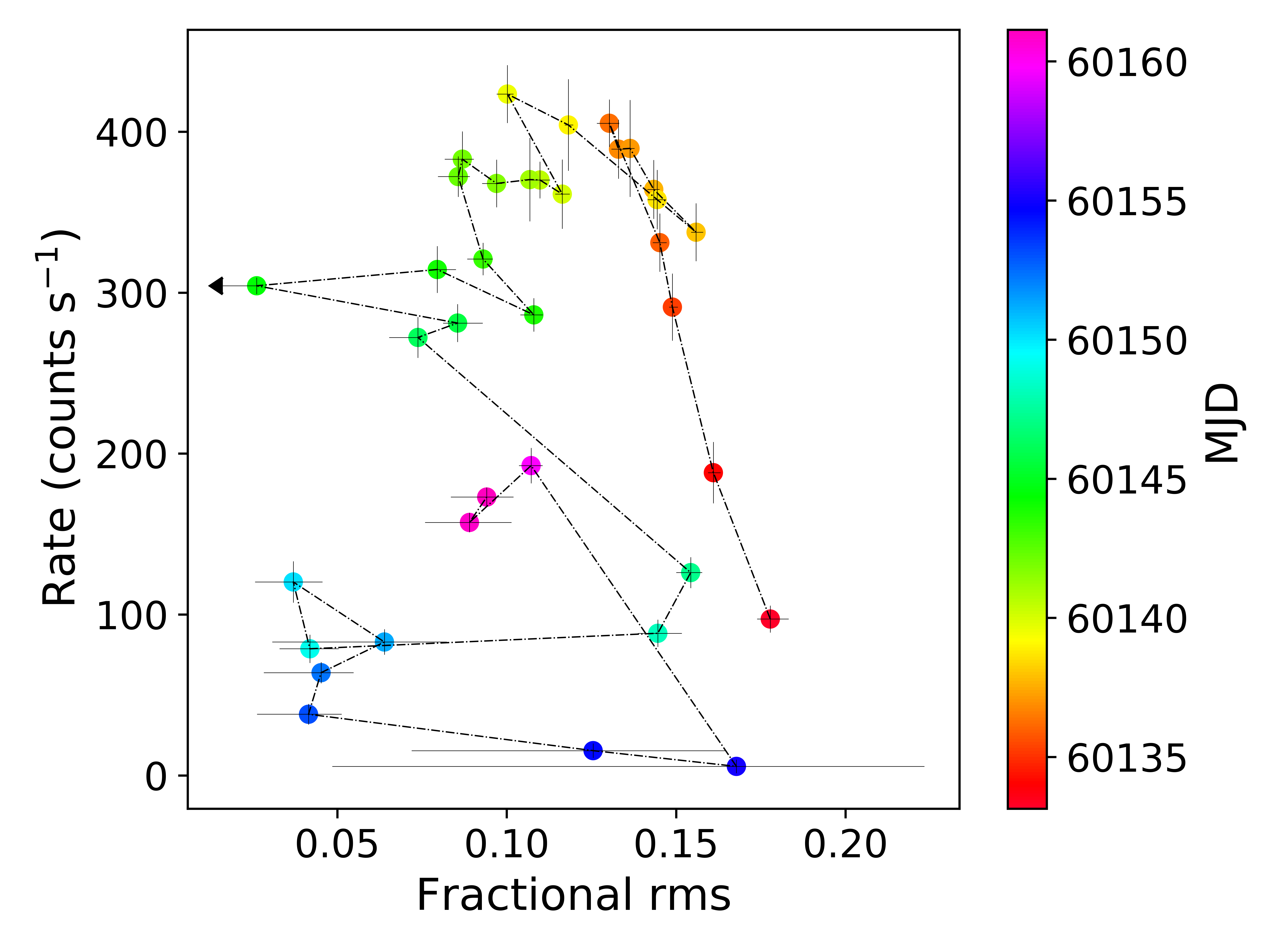}
    \caption{Left: Hardness-rms diagram (HRD) of \source{} during the 2023 outburst. The hardness ratio is the same as that defined in Figure \ref{fig:hidccd}. A positive correlation is apparent throughout the main outburst. The three observations during the reflare are systematically off the trend, having a low hardness ratio but an intermediate rms. Right: rms-intensity diagram (RID) of the source during the same period. Like the HID, here also a hysteresis loop is clearly detected.}
    \label{fig:hrdrid}
\end{figure*}

Figure \ref{fig:hidccd} shows the HID and CCD of \source{} during the outburst and the rebrightening after the outburst. In the HID, the hardness ratio is defined as the ratio of count rates in $2-10$ keV to $0.5-2$ keV energy bands, and the intensity in the ordinate represents the count rate in the $0.5-10$ keV range. The hard and soft colors in the CCD are defined as the ratios of count rates in $6-10$ keV to $4-6$ keV and $2-4$ keV to $0.5-2$ keV bands, respectively. The outburst begins with the source in a low luminosity hard state. The rapid rise in flux is associated with a similar decline in the hardness from $0.92 \pm 0.35$ ($1 \sigma$) to $0.18 \pm 0.01$ in the first four days. In the gradually declining plateau phase of the outburst that continues till MJD 60146.2, the hardness gradually decreases from $0.15 \pm 0.01$ to $0.12 \pm 0.01$. The sudden drop in flux marks an excursion of the source to the hard state after which it quickly returns to the soft state during the flare on MJD 60150.2. The final decay of the main outburst is associated with a corresponding increase in hardness up to the initial levels. The three observations during the bright reflare around MJD 60160 have a very low hardness ratio of $\approx 0.08$. The outburst of \source{} clearly traces a hysteresis loop on the HID with the initial hard-to-soft transition taking place at a higher count rate than the reverse transition from soft to hard state during the decay.

\subsubsection{X-ray Variability} 

The power spectra of \source{} take the shape of a band limited noise with a high frequency ($1-10$ Hz) cutoff and without any peaks. These were fitted by a single Lorentzian function where the centroid was consistent with 0 and thus, the width traced the cutoff frequency \citep{belloni02}. The total fractional rms was computed by taking the square root of the total area under the power spectrum in $0.1-10$ Hz. The dependence of the fractional rms on the hardness ratio and total count rate is depicted in figure \ref{fig:hrdrid}. In both the panels of figure \ref{fig:hrdrid}, the two lowest luminosity points at the beginning and end of the main outburst, i.e., on MJDs 60132.1 and 60158.4 (having ObsIDs: 6200840103 and 6634010120), are not included. This is because these observations were background dominated above 5 keV where the variability is expected to be high and, thus, the obtained rms is biased to low values from predominantly thermal X-rays. For similar reasons, the two observations on MJDs 60154.5 and 60155.1 have large uncertainty as these were dominated by background above 8 keV. The hardness-rms diagram (HRD, Figure \ref{fig:hrdrid}) shows a strong positive correlation. The Spearman rank-order correlation coefficient of the rms and hardness ratio is 0.84 and the p-value is $1.3 \times 10^{-10}$. If we consider the data only during the main outburst, leaving out the reflaring points, the correlation coefficient becomes 0.9 and the p-value decreases to $5.3 \times 10^{-13}$. The fractional rms - intensity diagram (RID) also shows a hysteresis loop in the counter-clock wise direction similar to the trends in other systems \citep{munozdarias11a, munozdarias14, alabarta20}. During the rise, the fractional rms starts from $17.8 \pm 0.4\%$ and decreases to $14.9 \pm 0.1 \%$ on MJD 60135.3. During the slow decay along the plateau phase the rms decreases to $7.4^{+0.1}_{-0.8}\% $ on MJD 60146.2. It continues to decrease to $4.2 \pm 1.2\% $ on MJD 60153.1 before making a short excursion to harder values of $15.4 \pm 0.4\%$ on MJD 60147.14 just before the short flare. Toward the end of the main outburst, the rms rises again to $16.8^{+11.9}_{-5.6}\%$ on MJD 60155.1, thus forming a hysteresis pattern. During the reflare after the end of the main outburst the rms remains at the 10\% level, even though the X-ray spectrum is soft at this time. It is remarkable that there appears to be a fairly tight correlation between hardness and fractional rms in the HRD, despite the source moving state and position in the HID, CCD and RID during this time. The data not following this correlation in the HRD are during the reflare, where the hardness is low while retaining $\sim 10$ per cent rms, and during the decay of the main outburst in the soft state, when the rms is slightly below the value predicted by the correlation.

With the help of the diagnostic diagrams (i.e., HID and RID) described above, along with the spectral properties to be discussed latter, we identified the approximate boundaries of the different states (Figure \ref{fig:lc}).  The observations during the reflare show a peculiar behavior where the variability ($10\%$) is of the intermediate state levels, while the hardness is quite low. 

\begin{figure*}
    \centering
    \includegraphics[scale=.55]{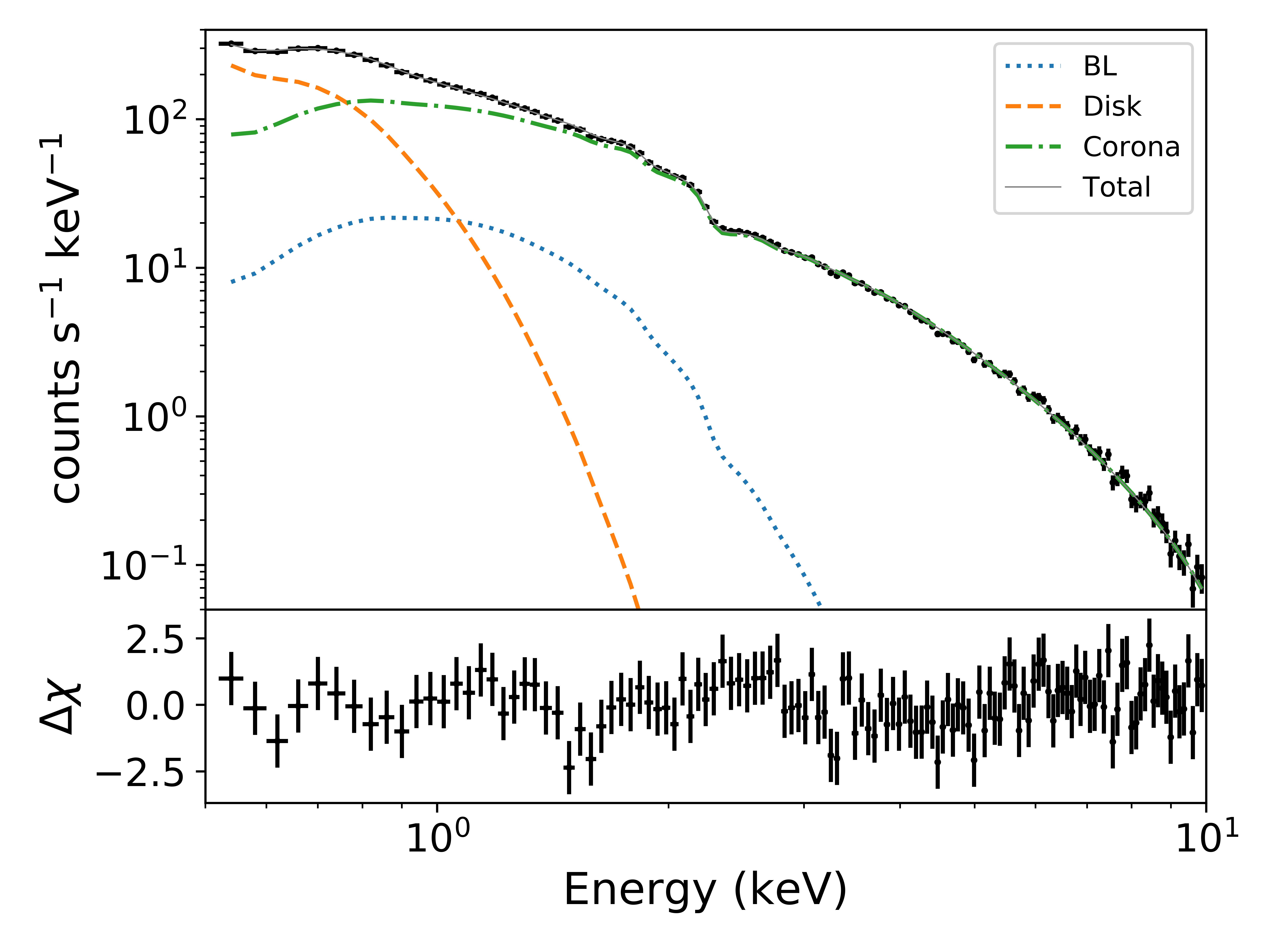}
    \includegraphics[scale=.55]{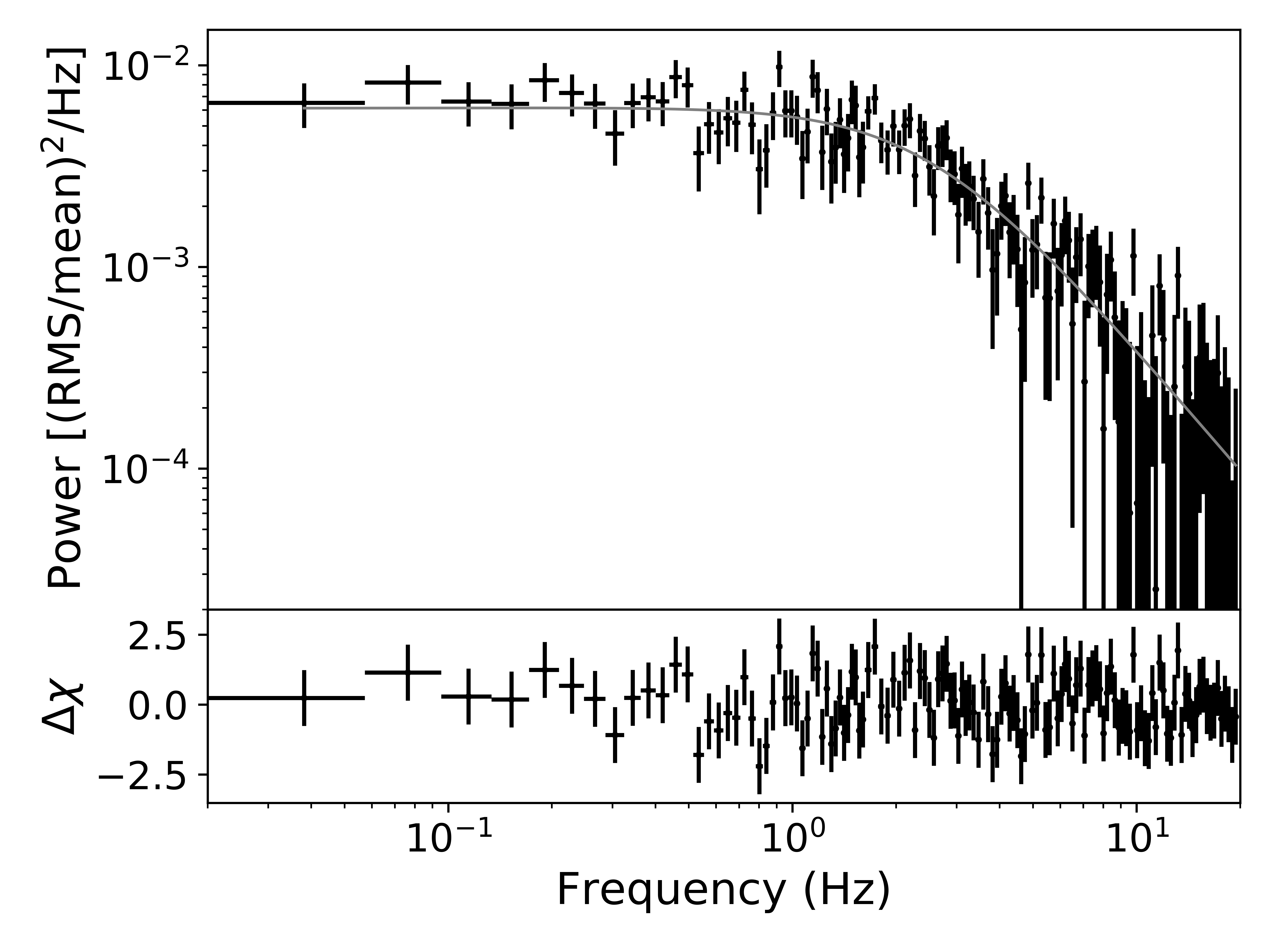}
    \caption{Left: A typical time-averaged spectrum with the individual model components observed on MJD 60138.68 (ObsId 6634010104). The dashed, dotted, and dot-dashed lines represent the disk, BL, and corona respectively. Right: Power spectrum of the same observation fitted by a single Lorentzian function.}
    \label{fig:specpds}
\end{figure*}

\subsubsection{X-ray Spectral Evolution} \label{sec:xrayspec}

As discussed in Section \ref{sec:intro}, the spectra of NS LMXBs generally require two thermal components originating from the BL and accretion disk. The detection of the two thermal components, however, depends on the sensitivity and energy range of the detectors. That is why many past studies were done with only a single thermal model. The spectra also require a power law component in the hard X-ray regime owing to inverse comptonization of the thermal photons in a hot corona. Accordingly, most spectra of \source{} required two thermal and a power law component (see the left panel of Figure \ref{fig:specpds} for a representative spectrum). For the thermal components, we used \dbb{} \citep{mitsuda84, makishima86} and \bb{}, and for the power law we used \nthc{} \citep{zdziarski96, zycki99}. We also used \tbabs{} to account for the absorption by the interstellar medium, wherein the photo-ionization cross-sections and abundances were adapted from \citet{wilms00}. The spectral fitting was carried out using the \texttt{XSPEC} package \citep{arnaud96}.  

Constraining the equivalent hydrogen column density, \nh{}, is tricky as it can behave similarly to the normalization of the thermal components, peaking at soft X-rays. Therefore, we fit all the \nicer{} spectra simultaneously, tying \nh{} across all the observations and letting other parameters that can change between the observations free. This results in a common value of \nh{} for the entire outburst and avoids any correlations with the thermal components. Residuals from the fit, however, showed structures in the $0.5-2$ keV range. While the calibration uncertainties in the Si K edge and Au M edge occur at 1.8 and 2.2 keV, respectively, the structures at lower energy were reminiscent of non-solar abundances of O K, Fe L, and Ne K edges \footnote{\url{https://iachec.org/wp-content/presentations/2020/NICER-IACHEC-Markwardt-2020a.pdf} }. Therefore, we replaced \tbabs{} with \tbfeo{} to let the abundances of Fe and O free\footnote{We also experimented with \tbvarabs{} to test if the abundance of Ne also varied. But, the best fitting Ne abundance remained consistent with the solar value.}. Thus, the final model used in the spectral analysis was \tbfeo{}*(\bb{} + \dbb{} + \nthc{}). The choice of seed photon temperature, \ktbb{}, for comptonization (\nthc{}) is not straightforward as both the inner-disk temperature, \tin{}, and the blackbody temperature of the BL, \kt{}, usually provide equally good fits. We tested both the model combinations and found that the best fitting \nh{} and abundances were almost the same. Leaving both Fe and O abundances in \tbfeo{} free produced a better fit compared to leaving either one of the Fe or O abundances free. However, we could not constrain the Fe abundance and only found a 3$\sigma$ upper limit of 0.13 times the solar abundance. The best fitting O abundance was $0.800^{+0.003}_{-0.016}$ times the solar value (1$\sigma$). \nh{} was constrained to $(2.616^{+0.021}_{-0.003}) \times 10^{21}$ cm$^{-2}$ (1$\sigma$) which is consistent with the estimate of \citet{jiminezibarra19} ($\gtrapprox 2.4 \times 10^{21}$ cm$^{-2}$) who used the equivalent width of NaI doublet to measure the line-of-sight extinction. This value is also consistent with the estimate by \citet[][$(2.4\pm 0.6) \times 10^{21}$ cm$^{-2}$]{shidatsu17} and slightly larger than the measurement by \citet[][$(1.3 \pm 0.9) \times 10^{21}$ cm$^{-2}$]{albayati21} from fitting the \swift{}/XRT and \nicer{} spectra during the discovery outburst of 2017. In the subsequent analyses pertaining to the individual observations, we fixed the best fitting parameters of \tbfeo{} obtained above.

% The model produced an excellent fit for all the spectra with a total \chisq{} of 4386.4 for 4213 degrees of freedom. BL seed gives 457§.85 for 4241 dof.

\begin{figure*}

    \centering
    \includegraphics[scale=.7]{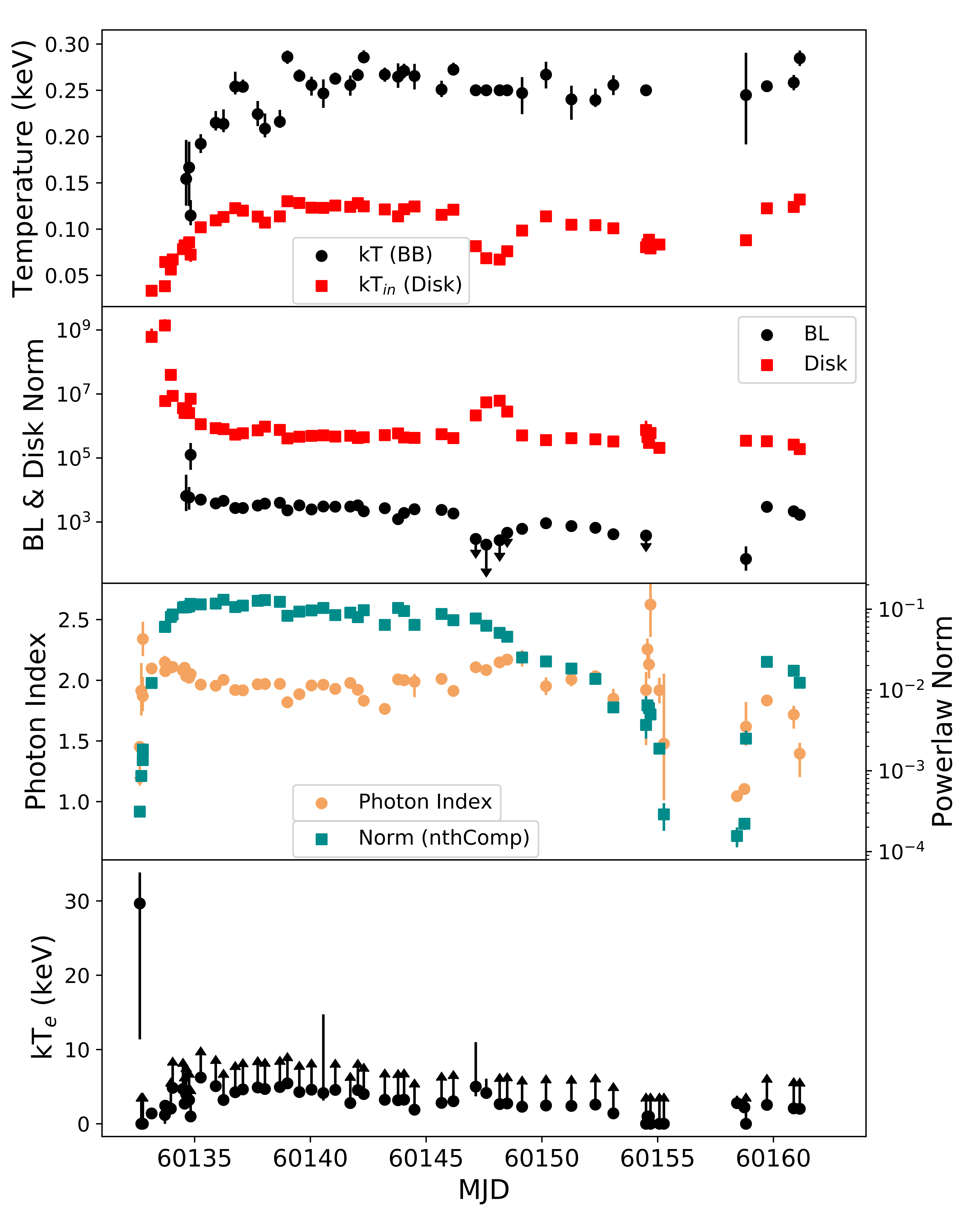}
    \caption{Temporal evolution of the best-fitting model parameters for all the X-ray spectra. The top two panels show the temperature and normalization of the two thermal components - disk and BL. The third panel depicts the evolution of the photon index and normalization of the comptonization component. The bottom panel shows the electron temperature. The lower limits are at $99\%$ significance.}
    \label{fig:bfp}
\end{figure*}

All three spectra in the first observation on MJD 60133 (6200840103) could be fitted by solely using \nthc{}. The source was very faint and the spectra were dominated by background above 5 keV. In the next observation on MJD 60134 (6200840104), the spectra for all the segments required only one thermal component, i.e., a multi-color blackbody from the disk, besides the comptonization component. It was only from the fourth segment of the third observation (6200840105; MJD 60135) that an additional single-temperature blackbody corresponding to the BL component became necessary to fit the spectrum. Adding the BL component improved the $\chi^2$ by 19 for 2 degrees of freedom and resulted in an F-test probability of $4.8 \times 10^{-4}$. In the subsequent observation, the improvement in $\chi^2$ was 57 for 2 degrees of freedom and the F-test probability came out to be $7.2 \times 10^{-11}$. As the source transitioned to the intermediate state, the fits with a single thermal component were not even formally acceptable with $\chi^{2}_{\nu} \gtrsim 2$. The BL again became unconstrained in the two epochs on  MJD 60147 and 60148 following the intermediate state, when there was a drop in the count rate. Similarly, toward the end of the main outburst and the beginning of the reflare, the BL component was also not required. Apart from the actual dimming of the BL component, the background domination above 5 keV in the spectra made it difficult to constrain an extra thermal component. There exists a degeneracy in the source of seed photons for comptonization in NS LMXBs as both the disk and BL can contribute the same. For 4U 1608--52, \citet{armaspadilla17} suggested that the BL most likely acts as the seed photon source during soft state, whereas in the hard state both solutions were equally favored. Incidentally, as noted above, for \source{} the BL was either unconstrained or not required during the hard states\footnote{We note that \citet{armaspadilla17} used \textit{Suzaku} data that covered $0.8-30$ keV. This gave them better control over the comptonization component.}. Thus, in our analysis, we let the BL be the seed photon source in the soft and intermediate states and the disk in the hard states. We note that the best-fitting parameters and their general trends remain very similar in both the cases.

 \begin{figure*}
     \centering
     \includegraphics[scale=0.55]{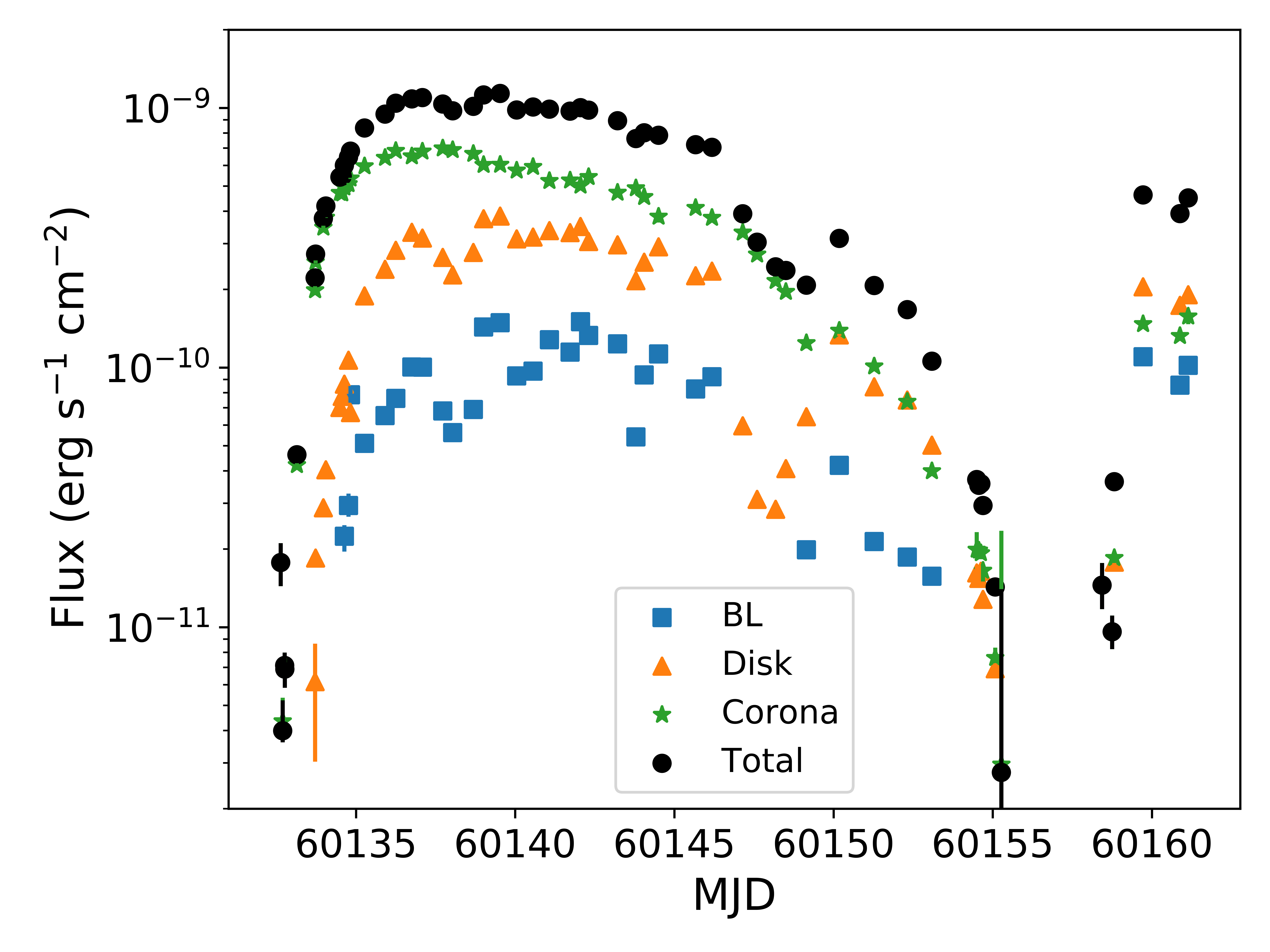}
     \includegraphics[scale=0.55]{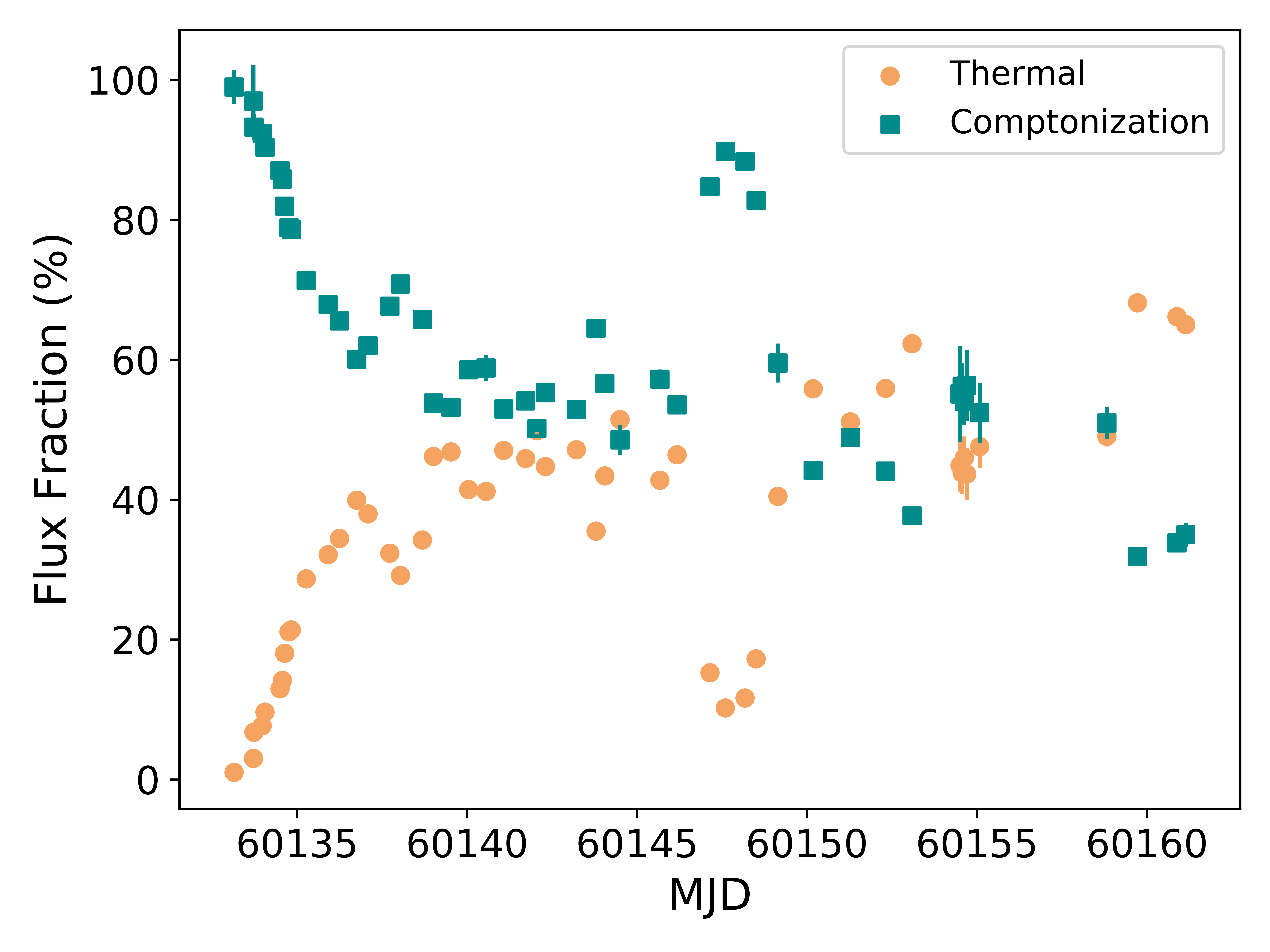}
     \caption{Time evolution of the fluxes of the individual model components (left), and the fraction of thermal and comptonization flux (right).}
     \label{fig:flux}
 \end{figure*}
 
The best fitting parameters of the individual components of the model are plotted in figure \ref{fig:bfp}. The first and second panels contain the temperature and normalization of the two thermal components. \tin{} starts from about 0.03 keV and stabilizes at $\approx 0.11$ keV for the rest of the outburst, except at the end of the intermediate state, around MJD 60147 where it dips to $\sim 0.07$ keV. Similarly, \kt{} starts from $0.15$ keV and gradually rises to $\approx 0.25$ keV. It is quite remarkable that these values of the respective temperatures are much lower than those typically observed in other NS LMXBs \citep{lin07,armaspadilla17,banerjee24}. The normalization of the two thermal components follows a reverse trend. The \norm{} of \dbb{} starts from $6.05 \times 10^{8}$ and gradually decreases to $\approx 5 \times 10^5$ while the \norm{} of \bb{} decreases gradually from $6.49 \times 10^3$ to $3.76 \times 10^{2}$ near the end of the main outburst. Following the end of the intermediate state (MJD 60147), the \norm{} of \dbb{} increases by about an order of magnitude. The BL component, however, could not be constrained. Therefore, we fixed \kt{} to 0.25 keV and obtained upper limits for its \norm{}. Figure \ref{fig:bfp} shows the evolution of the \norm{} and \gama{} of \nthc{}. The \norm{} moves from $3.1 \times 10^{-4}$ to $\sim 0.1$ during the rise and then continues to decay gradually. At the end of the main outburst, the \norm{} decays to $2.89 \times 10^{-4}$. During the reflare, unlike the thermal components which rise almost to the same level as the main outburst, the \nthc{} \norm{} remains about an order of magnitude below at $\sim 0.02$. Apart from the first observation, where it could not be constrained, \gama{} remains quite stable for the entirety of the main outburst at $\approx 2$. It rises to slightly higher values of 2.5, albeit with larger uncertainty, during the soft state just before the end of the main outburst. During the reflare, \gama{} varies between $1.3-1.8$. Finally, in the last panel we show the electron temperature \kte{}. Owing to the limited energy range of \nicer{}, \kte{} could not be constrained for the majority of the observations. It was only for a few observations during the hard states (i.e., around MJD 60132, 60147, and 60158) that a significant detection could be made. For all other cases, we report a $99\%$ lower limit that varied between $0-5$ keV.

The low value of \gama{} during the reflares is consistent with the hard/intermediate state. This explains the higher ($\sim 10\%$) variability detected during these epochs. The soft color, on the other hand, could be due to something unusual happening at lower energies. It could be due to variable absorption, an extra component, or simply depend on the choice of the low energy band wherein the relative contribution of the power law is systematically low. Indeed, we find that the shape of the spectrum during the reflares has a larger thermal bump. The fit statistics during these three epochs are also much worse compared to rest of the observations. We tried freeing up the \nh{} and found the best fitting \nh{} increase by a factor of $1.3-1.7$. Consequently, the normalization of both the thermal components also increased, providing a significant improvement in fit. 

In figure \ref{fig:flux}, we show the evolution of the unabsorbed flux in the 0.5$-$10 keV range from the individual components of the spectra. The fluxes were calculated by attaching the convolution model \texttt{cflux} to the individual components. The comptonization component, \nthc{}, dominates the total flux for the major part (i.e., from the beginning till MJD 60150.18) of the main outburst. In fact, it turns out that the fraction of compton flux never falls below $30\%$ in the entire outburst. The lowest level it reaches is during the reflare when it falls to $\approx 33\%$. Consequently, the fraction of total thermal flux, i.e., disk and BL included, never increases above $70\%$, the highest contribution being during the reflare at $\approx 67\%$. Even during the soft state between MJD 60150 to 60153, when the fractional rms is $\sim 5\%$, the fraction of thermal flux lies between $50-60\%$.

\section{Discussion}

The evolution of \source{} on the X-ray CCD and HID (Figure \ref{fig:hidccd}) and its spectral and timing properties resemble a typical atoll source \citep{hasinger89}. This implies that it remains at sub-Eddington or low accretion rate levels, i.e., $0.01 - 0.5 L_{Edd}$. The source follows a hysteresis track in the HID and RID making full transitions from the hard to soft state through an intermediate state, and then back to the hard state at a lower luminosity \citep[][; Figure \ref{fig:hrdrid}]{munozdarias14}. 

% spectral properties 
The spectral decomposition of NS LMXBs is more complicated than their BH counterparts, owing to the presence of a hard surface and magnetic fields that are anchored on it. These extra structures result in an additional thermal component from the BL or the disk, which might be truncated at a larger radius defined by the magnetic field strength. Accordingly, most of the spectra of \source{} required three components for a satisfactory fit. The disk temperature was always found to be lower than the BL \citep{mitsuda89, lin07}. The BL was not needed at the very beginning and end of the main outburst. Just after the intermediate state, the BL component could also not be constrained (Figure \ref{fig:bfp}). During these hard state observations with a single thermal component, the presence of a disk was not only statistically favored over the BL emission, but also the best fitting normalization and disk temperature were more consistent with the evolution of the disk. For observations requiring both disk and BL emission, there was degeneracy in the source of seed photons for comptonization. In this work, we set the seed photons to BL in the soft and intermediate states and to the disk in the hard states, where either the BL was unconstrained or undetected \citep[e.g.][]{armaspadilla17}.

\begin{figure*}
    \centering
    \includegraphics[scale=0.55]{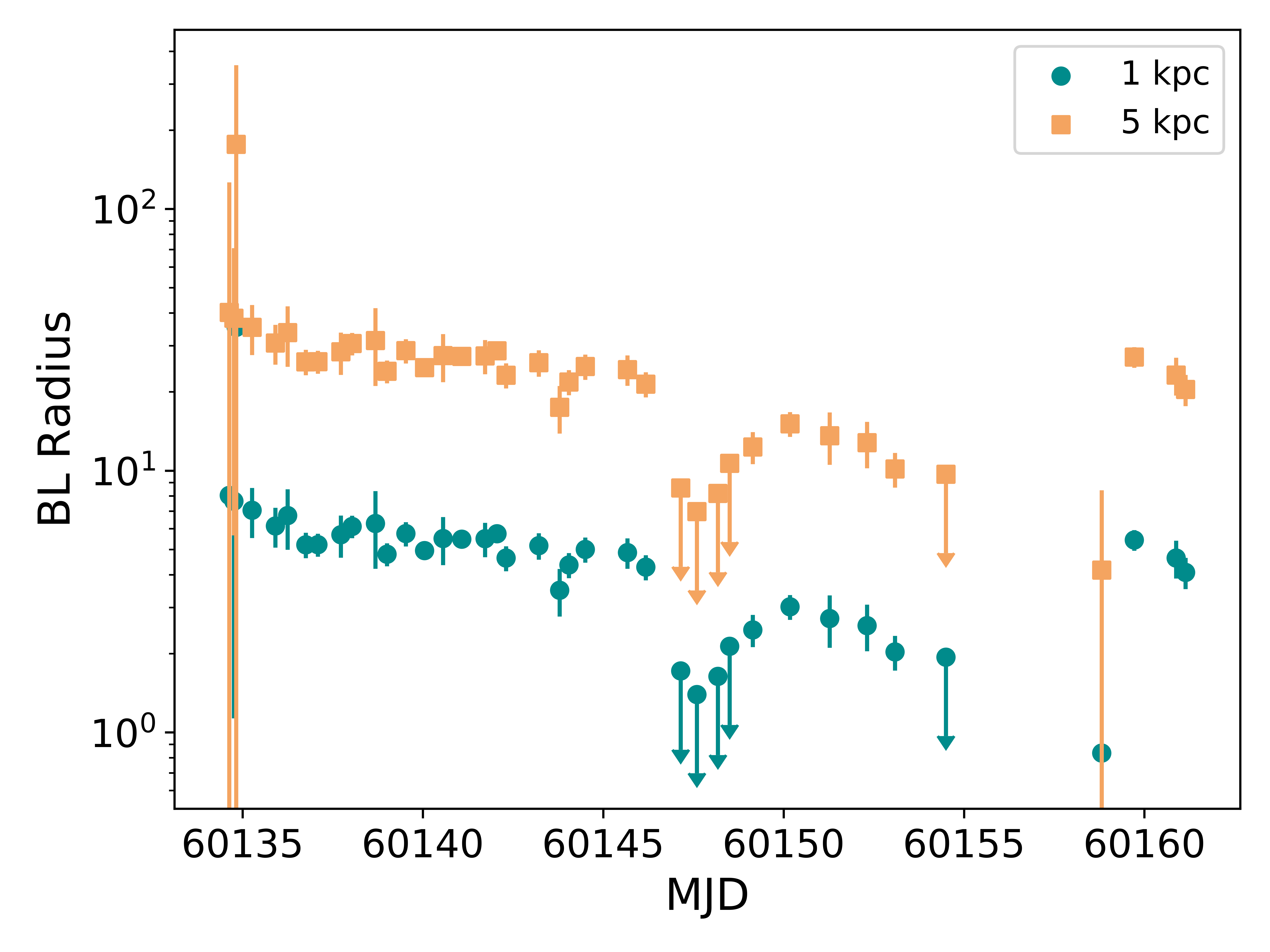}
    \includegraphics[scale=0.55]{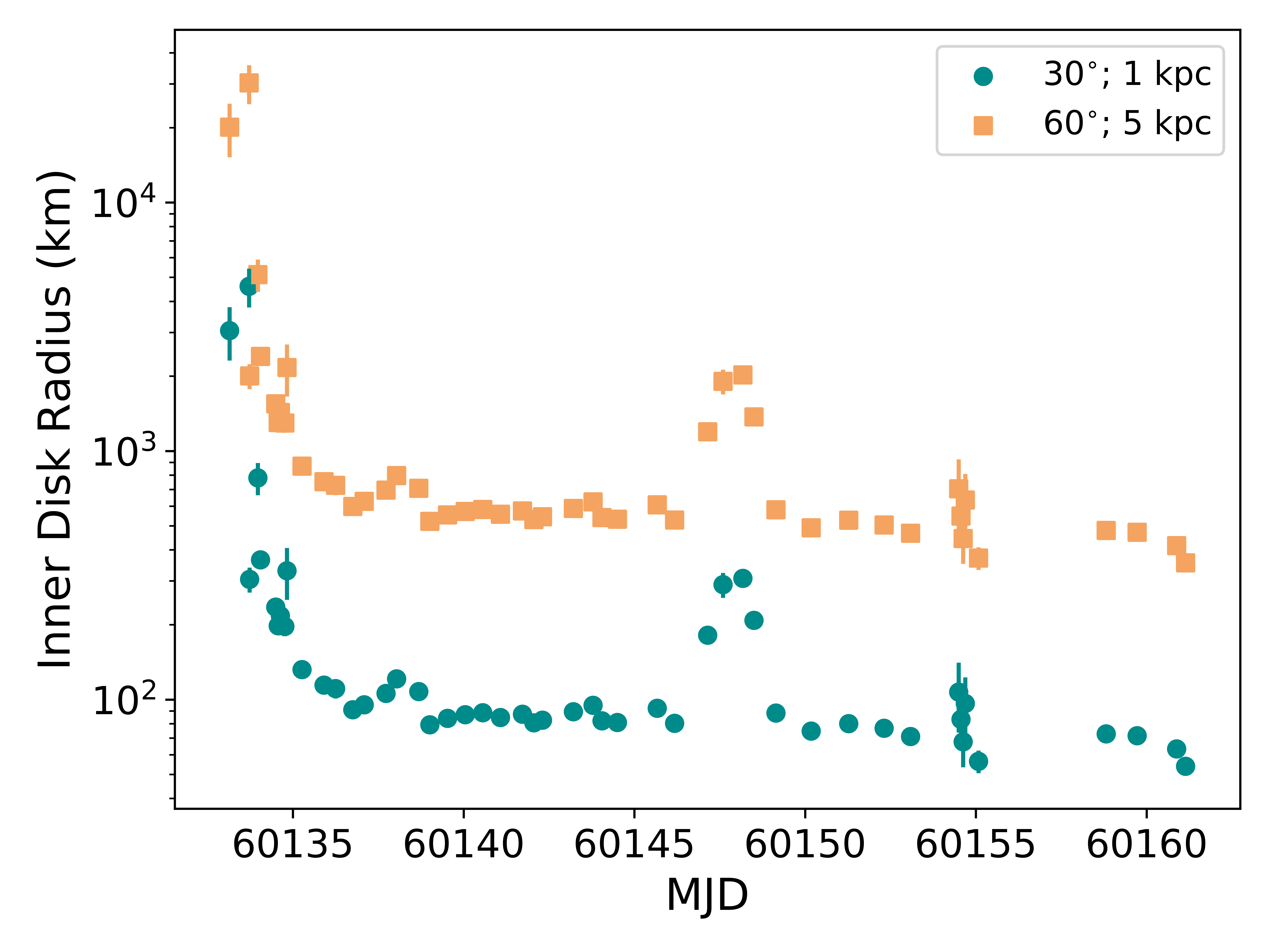}
    \caption{Values of the BL radius (left) and inner-disk radius (right) obtained from the best-fitting normalization of the respective thermal components.}
    \label{fig:radius}
\end{figure*}

The normalization of the two thermal components can be used to estimate the radius of the BL and inner disk. While the former depends on the distance to the source, the latter depends both on the distance and inclination. Both these quantities are unknown for our source, hence we can only calculate a range of best possible values. From the X-ray/optical correlations of \source{}, \citet{jiminezibarra19} suggested that the distance of the source has to be low ($1-5$ kpc) in order for the compact object to be a NS. Within $1-5$ kpc, the radius of the blackbody emitting surface varies between $2-8$ km for 1 kpc and between $10-40$ km for 5 kpc (Figure \ref{fig:radius}). The radius of a typical NS $\approx 10$ km, thus favoring a higher value for the distance. But if the emission is due to some residual accretion onto the NS surface, the BL can be quite small \citep[e.g.][]{lin07} and the distance large. This range of distances is well within the weaker constraint of $<12.4$ kpc set by \citet{albayati21} from Type I bursts. For the inner radius of the disk, we need two correction factors, one owing to the difference between color and effective temperature \citep[$\kappa=1.7$;][]{shimura95} and another that adjusts the flux for a non-zero torque at the inner boundary \citep[$\zeta=0.4$;][]{kubota98}. We calculated the radius for two extreme cases - one with low values of distance and inclination (i.e., 1 kpc and $30^\circ$), and the other with high values (i.e., 5 kpc and $60^\circ$). The evolution of the disk inner radius (Figure \ref{fig:radius}) shows that the disk is truncated at a large distance at the beginning of the outburst and then move inward rapidly, decreasing by about a factor of 40 in a matter of three days. The inner radius remains stable at this level throughout the intermediate state, then throughout the soft state and also during the reflares. Therefore, it can be considered as the last stable orbit of the disk. For the large distance - high inclination case, this radius is $\sim 500$ km (or $\approx~ 240~ R_g$) and for the small distance - low inclination case it is $\sim 80$ km (or $\approx~ 40~ R_g$). The former value is much larger than the disk radius typically inferred in both pulsating and non-pulsating NS systems, while the latter is just in the ballpark, although still a factor of a few higher \citep{cackett10,millerjm11b,ludlam19}. Assuming that the disk halts at the magnetospheric, or Alfv\'en, radius, we estimate the magnetic field strength with the relation \citep{ibragimov09,cackett09,degenaar17}:

$$
\begin{aligned}
    B = 1.2 \times 10^{5} ~k_A^{-7/4} ~ \left(\frac{R_{in}}{R_g}\right)^{7/4} \left(\frac{M_\text{\tiny NS}}{1.4~ M_\odot} \right)^2 ~ \frac{D}{5~ \text{kpc}} \\
    \left(\frac{R_\text{\tiny NS}}{10^6 ~ \text{cm}}\right)^{-3} \times \left(\frac{f_{ang}}{\eta} \frac{F_{bol}}{10^{-9}~ \text{erg s$^{-1}$ cm$^{-2}$}}\right)^{1/2} ~\text{G},
\end{aligned}
$$

where $B$ is the magnetic field, $f_{ang}$ is the anisotropy correction factor \citep[$1.1-1.5$;][]{ibragimov09}, and $k_A$ is the spherical to disk geometry correction factor \citep[$0.5-1.1$;][]{long05, psaltis99b, kluzniak07}. The peak bolometric flux, $F_{bol}$, obtained by expanding the response of the \nicer{} spectra to $0.1-200$ keV range, was found to be $3.7 \times 10^{-9}$ \fluxcgs{}. This bolometric flux will have some uncertainty, as the cutoff energy of the power law may lie somewhere between $40-100$ keV. However, since the spectrum during the peak is quite steep ($\Gamma \gtrsim 2$)the bulk of the flux is emitted at energies below $\sim 20$ keV, i.e. $\gtrsim 80\%$. The uncertainty owing to the range of $f_{ang}$ and $k_A$ is much larger than the uncertainty due to bolometric flux estimation. Assuming the  efficiency of accretion, $\eta = 0.1$,  $M_\text{\tiny NS} = 1.4~ M_\odot$, and $R_\text{\tiny NS} = 10$ km, we get $B \approx 0.8-3.5 \times 10^8$ G for $D =$ 1 kpc and $B \approx 9.4-43.4 \times 10^{9}$ G for $D=$ 5 kpc. The latter estimate is  much higher than those measured from typical atolls, which have a $B$ field of $\approx 10^8$ G \citep[e.g.][]{ludlam19}. While it may be tempting to favor a small distance to the source, that will make the peak luminosity very low. For a distance of 1 kpc the peak luminosity will be $\sim 1.2 \times 10^{35}$ erg s$^{-1}$ (or $\sim 0.0008 ~ L_{Edd}$), while for 5 kpc it will be $\sim 3.0 \times 10^{36}$ erg s$^{-1}$ (or $\sim 0.02 ~ L_{Edd}$). The luminosities at smaller distances are too low, especially given that the source has made a clear transition to the soft state which is expected to occur at luminosities $> 1-4\% ~ L_{Edd}$ \citep{maccarone03}.   

On MJD 60147, \source{} underwent a sharp and short transition to the intermediate or even a hard state (Figure \ref{fig:hidccd}), before returning to the soft state. During this hard excursion, the thermal flux decreased by about an order of magnitude, while the power law flux decreased only slightly (Figure \ref{fig:flux}). The decrease in flux is also reflected in a corresponding increase in the truncation radius of the disk (Figure \ref{fig:radius}). There is no clear explanation for these types of state transitions, which could either be due to changing disk properties, such as the accretion rate \citep{maccarone03}, or due to the propeller effect \citep{asai13}. A strong propeller regime is unlikely, as this would lead to very little accretion onto the NS surface, the signature of which we clearly detect. Accretion in \source{} could then be occurring in a trapped-disk \citep{dangelo12} or weak propeller regime \citep{ustyugova06}. Under both these regimes, a considerable amount of matter can accrete onto the NS surface and the source displays weak outflows. Furthermore, in the weak propeller regime, the magnetospheric radius is expected to be smaller than the co-rotation radius ($R_{co}$). So an estimate of the magnetospheric radius inferred above acts as a lower limit on $R_{co}$. This, in turn, can give us a lower limit on the spin period of the NS from the relation $R_{co} = (GM_{NS}P^2/4{\pi}^2)^{1/3}$ where $G$ is the gravitational constant and $P$ is the spin period. An $R_{co}$ varying between $40-500$ $R_g$, for distances within $1-5$ kpc, results in $P \gtrsim 10.4-162.9$ ms. The possible values of period are larger than many NS LMXBs with known pulsations and burst oscillations \citep[$1.6-10$ ms;][]{patruno10}, though they are not that atypical\footnote{\url{http://www.iasfbo.inaf.it/~mauro/pulsar_list.html}}.

NS LMXBs do not show very low variability (i.e., rms $< 1\%$) as there is always asignificant contribution from the comptonization component, which probably gets trapped in the magnetic field and does not get completely depleted. Also, owing to the continuous BL emission, the total rms gets diluted and does not reach very high values ($> 30\%$). Consequently, both these effects result in a rather narrow dynamic range for the variability compared to BH XRBs. The fractional rms of \source{} varied between $5\% - 17\%$ agreeing with this principle. The rms also shows a strong positive correlation with the hardness ratio (Figure \ref{fig:hrdrid}) consistent with an origin of the variability in the comptonizing medium \citep{munozdarias14, alabarta20}. The characteristic frequency, \numax{}, of the power spectrum represents the Fourier frequency at which maximum power is dissipated \citep{belloni02}. For narrow components, such as QPOs, \numax{} is almost equal to the centroid frequency, $\nu_0$. However, as the component becomes broader, \numax{} approaches the value of the width of the Lorentzian \citep{belloni02}. In power spectra that consist of a band-limited noise (BLN), like those observed in \source{} (Figure \ref{fig:specpds}), \numax{} traces the break frequency. 

\begin{figure}
     \centering
     \includegraphics[scale=0.5]{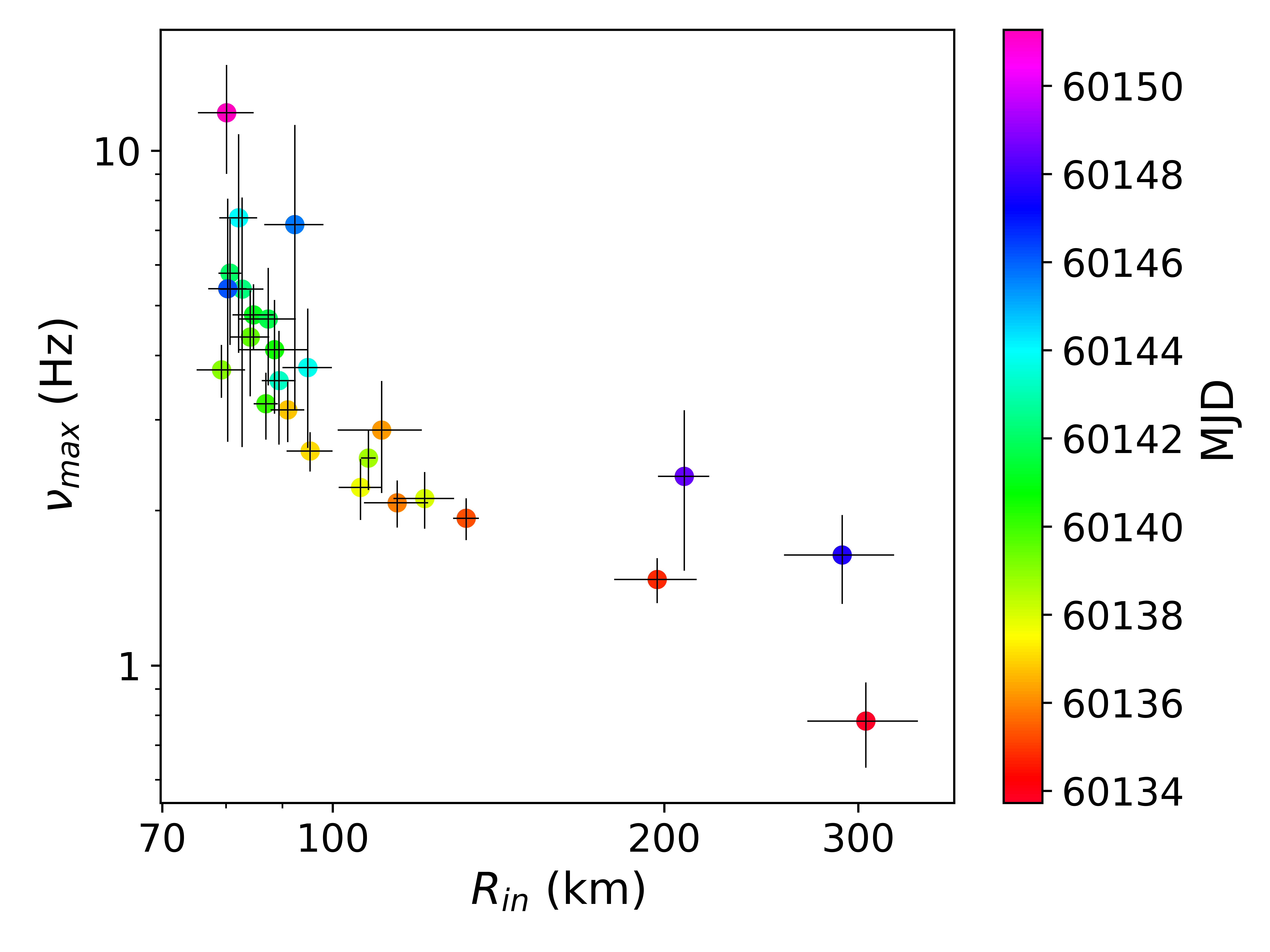}
     \caption{The correlation between the characteristic frequency of the power spectra with the disk inner radius from the beginning until the end of the excursion in the hard state.}
     \label{fig:numax}
\end{figure}

The break frequency of the BLN is known to be strongly correlated with the QPO frequency in both BH and NS XRBs, including millisecond pulsars \citep{wijnands99}. In the bursting atoll 4U 1728--34, the BLN was seen to evolve into a QPO with state transition \citep{disalvo01a}. \citet{maccarone11}, using a bispectrum technique, showed that the QPOs are coupled with the broadband noise components in GRS 1915$+$105 rather than being generated independently. Although the origins of these timing features are still debated, the associated frequencies are generally believed to trace the size of the emitting region \citep[e.g.][]{ingram09, garain14, marcel20}. The viscous frequency of this comptonizing medium is anti-correlated with the outer radius of the flow \citep[e.g.]{narayan12,marcel18}. If the variability is assumed to arise from propagation of fluctuations in the accretion rate, the transition radius from a stable to turbulent flow is traced by the low frequency component of the power spectrum \citep{churazov01, kawamura22}. The characteristic frequency of the BLN in \source{} is found to be strongly anti-correlated with the disk inner radius, consistent with the prediction of the above models (Figure \ref{fig:numax}). The Spearman correlation coefficient between the two independently measured quantities is -0.87 and the p-value is $\sim 5 \times 10^{-9}$. This suggests that the size of the corona significantly shrinks as the source transitions from a hard to intermediate state. The size even increases for a short duration before the decay, when the source makes an excursion to the hard state. 

While the spectral fits provide information about the evolution of the inner radius of the accretion disk and the size of the BL (Figure \ref{fig:radius}), the fast variability properties inform us about the evolution of the size of the corona (Figure \ref{fig:numax}). Combining this information, we suggest a possible scenario for the evolution of the disk/corona geometry during the various stages of the outburst. At the beginning, when the source is in the hard state, the corona fills up the entire space between the NS and the inner edge of the truncated accretion disk. With the rise in accretion rate, the disk moves inward and with it the corona contracts. This contraction of the comptonizing medium occurs rapidly during the hard state (the first few days) and then gradually in the intermediate state, when the disk has already stabilized at the last orbit. From a multi-wavelength study, signatures of jet emission has been detected during the intermediate state (Rout et al. 2025, in preparation). During this period, perhaps the base of the jet acts as the comptonizing medium and the gradual shrinking of it is due to the emptying of matter into the outflow \citep[][]{beloborodov99,fender04b}. The possible change in the geometry of the corona is also in agreement with our consideration of the BL acting as the source of seed photons due to its proximity with the jet base compared to the disk. Some recent studies using spectral-timing methods have suggested a changing coronal geometry with spectral states \citep[e.g.][]{mendez22, garciaf22}. In the subsequent phase of hard excursion, the disk recedes by about a few hundred km, the BL becomes unconstrained, but the jet, although weak, is still present (Rout et al. 2024, in preparation). The disk then returns to the last stable orbit for a short while when the source transitions to the soft state before the end of the main outburst.

\section{Summary}

In this work, we present the results of a comprehensive X-ray spectral and timing analysis of the NS transient \source{} during its latest outburst beginning July 2023. \source{} follows the general trends in spectral and variability properties of LMXBs along with having some peculiarities. It traces a hysteresis track in the HID and RID, marking full spectral state transitions. The results from X-ray spectroscopy are consistent with a truncated disk in the hard states. The values of inner disk radius for a large distance (5 kpc) result in a strong magnetic field ($\sim 10^{10}$ G) if the disk is assumed to halt at the magnetospheric radius. On the other hand, a weaker magnetic field ($\sim 10^8$ G) can be obtained from a small distance (1 kpc), but that results in a very low peak luminosity ($\sim 0.07~\% ~L_{Edd}$). The evolution of the characteristic frequencies of the power spectrum suggests a drop in the size of corona as the source transitions from a hard to an intermediate state. We speculate that a change in coronal geometry from horizontal to vertical occurs as the source transitions from a hard to intermediate state, which could be a cause of the change in the seed photon source. During the reflare after the end of the main outburst, there are indications of variable absorption in the soft X-rays, the origin of which is unclear.

\section{Acknowledgement}

The authors acknowledge the constructive feedback from the referee. This research is based upon work supported by Tamkeen under the NYU Abu Dhabi Research Institute grant CASS. This research has made use of data and/or software provided by the High Energy Astrophysics Science Archive Research Center (HEASARC), which is a service of the Astrophysics Science Division at NASA/GSFC. SKR acknowledges K Sriram and Ranjeev Misra for useful discussions. TMD acknowledges support by the Spanish Ministry of Science via the Plan de Generacion de conocimiento PID2021-124879NB-I00. JH acknowledges support through NASA grant 80NSSC23K1659. MAP acknowledges support through the Ramón y Cajal grant RYC2022-035388-I, funded by MCIU/AEI/10.13039/501100011033 and FSE$+$.

% \newpage
\bibliography{references}{}
\bibliographystyle{aasjournal}

\end{document}